\documentclass[prd,twocolumn,groupedaddress,showpacs,nofootinbib]{revtex4}
\usepackage{graphicx}
\usepackage{dcolumn}
\usepackage{bm}
\usepackage{amssymb}
\usepackage{mathrsfs}
\usepackage{amsmath}
\usepackage{epsfig}
\usepackage[dvips]{color}
\usepackage{hhline}

\begin{document}

\title{Double and Linear Seesaw from Left-Right and Peccei-Quinn Symmetry Breaking}

\author{Pei-Hong Gu}
\email{peihong.gu@mpi-hd.mpg.de}

\affiliation{Max-Planck-Institut f\"{u}r Kernphysik, Saupfercheckweg
1, 69117 Heidelberg, Germany}

\begin{abstract}

In the left-right symmetric models, we can take the left-right
symmetry to be the charge-conjugation and then impose a global
symmetry under which the left- and right-handed fermion doublets
carry equal but opposite charges. Consequently, we may introduce two
Higgs bi-doublets to give the desired fermion mass spectrum. The
global symmetry is identified to a Peccei-Quinn symmetry. We can
introduce a complex scalar singlet to break the global Peccei-Quinn
symmetry at a high scale. This symmetry breaking is also responsible
for generating the heavy Majorana masses of the fermion singlets
which have Yukawa couplings with the lepton and Higgs doublets. In
this context, we can realize the double and linear seesaw to
naturally explain the small neutrino masses. Our scenario can be
embedded in the $SO(10)$ grand unification theories.

\end{abstract}

\pacs{12.60.Cn, 12.60.Fr, 14.80.Va, 98.80.Cq}

\maketitle

\section{Introduction}

In the left-right symmetric models \cite{ps1974} based on the gauge
group $SU(3)_{c}^{}\times SU(2)_{L}^{}\times SU(2)_{R}^{}\times
U(1)_{B-L}^{}$, the left-handed fermions are $SU(2)_L^{}$ doublets
as they are in the $SU(3)_{c}^{}\times SU(2)_{L}^{}\times
U(1)_{Y}^{}$ standard model (SM) while the right-handed fermions
(the SM right-handed fermions plus the right-handed neutrinos) are
placed in $SU(2)_R^{}$ doublets. Usually, we need a Higgs bi-doublet
to construct the Yukawa interactions between the left- and
right-handed fermion doublets for generating the Dirac masses of the
SM fermions. As for the small neutrino masses, they are given by the
seesaw \cite{minkowski1977,mw1980}, which also accommodates the
leptogenesis \cite{fy1986} to explain the matter-antimatter
asymmetry in the universe. For example, we can extend the original
left-right symmetric model \cite{ps1974}, where the Higgs scalars
include one bi-doublet and two doublets, by introducing three
fermion singlets with a Majorana mass term. In the presence of the
Yukawa couplings of the fermion singlets to the lepton and Higgs
doublets, we can obtain the double/inverse
\cite{mohapatra1986,mv1986} and linear \cite{barr2003} seesaw for
generating the small neutrino masses to revive the original
left-right symmetric model \cite{gs2010}.

The left-right symmetry can be the parity or the charge-conjugation.
In the case that the left-right symmetry is the charge-conjugation,
we can impose a global symmetry under which the left- and
right-handed fermion doublets carry equal but opposite charges. In
this context, the global symmetry should be explicitly broken to
generate the neutrino and charged fermion masses. For example
\cite{gl2010}, the global symmetry breaking can naturally make the
fermion singlets, which have the Yukawa couplings with the fermion
and Higgs doublets, to obtain the heavy masses for realizing the
universal seesaw \cite{berezhiani1983,cm1987,bms2003,gu2010}. Since
the global symmetry is mediated to the SM quarks, it can be
identified to the Peccei-Quinn \cite{pq1977} (PQ) symmetry, which
predicts the axion \cite{pq1977,weinberg1978} to solve the strong CP
problem.

In this paper we will revive the original left-right symmetric model
by the PQ symmetry breaking. Specifically we will take the
left-right symmetry to be the charge-conjugation and then assign the
equal but opposite PQ charges for the left- and right-handed fermion
doublets. We then need two Higgs bi-doublets to generate the desired
fermion masses. This structure definitely will induce the
Peccei-Quinn-Weinberg-Wilczek (PQWW) axion
\cite{pq1977,weinberg1978}. To make the axion invisible, we will
further introduce a complex scalar singlet for the PQ symmetry
breaking, as did in the Kim-Shifman-Vainshtein-Zakharov
\cite{kim1979} (KSVZ) model and the
Dine-Fischler-Srednicki-Zhitnitsky \cite{dfs1981} (DFSZ) model.
After the PQ symmetry breaking, the fermion singlets can obtain
their heavy Majorana masses to realize the double and linear seesaw.
Our scenario can be embedded in the $SO(10)$ grand unification
theories (GUTs).

\section{The model}

The scalar fields include two Higgs bi-doublets,
\begin{eqnarray}
\begin{array}{c}
\phi_1^{}(\textbf{1},\textbf{2},\textbf{2}^\ast_{},0)=\left[\begin{array}{cc}\phi_{11}^0&\phi_{12}^+\\
[2mm]\phi_{11}^-&\phi_{12}^0\end{array}\right]\,,\\
[7mm]
\phi_2^{}(\textbf{1},\textbf{2},\textbf{2}^\ast_{},0)=\left[\begin{array}{cc}\phi_{21}^0&\phi_{22}^+\\
[2mm]\phi_{21}^-&\phi_{22}^0\end{array}\right]\,,
\end{array}
\end{eqnarray}
two Higgs doublets,
\begin{eqnarray}
\begin{array}{c}
\chi_L^{}(\textbf{1},\textbf{2},\textbf{1},-1)=\left[\begin{array}{l}\chi_L^0\\
[2mm] \chi_L^-\end{array}\right]\,,\\
[7mm]
\chi_R^{}(\textbf{1},\textbf{1},\textbf{2},-1)=\left[\begin{array}{l}\chi_R^0\\
[2mm] \chi_R^-\end{array}\right]\,,
\end{array}
\end{eqnarray}
and one complex singlet,
\begin{eqnarray}
\sigma(\textbf{1},\textbf{1},\textbf{1},0)\,.
\end{eqnarray}
In the fermion sector, there are three neutral fermion singlets,
\begin{eqnarray}
S_R^{}(\textbf{1},\textbf{1},\textbf{1},0)\,,
\end{eqnarray}
besides three generations of quark and lepton doublets,
\begin{eqnarray}
\begin{array}{l}q_{L}^{}(\textbf{3},\textbf{2},\textbf{1},~\,\frac{1}{3})=\left[\begin{array}{c}
u_L^{}\\
d_L^{}
\end{array}\right]\,,\\
[5mm]
l_{L}^{}(\textbf{1},\textbf{2},\textbf{1},-1)=\left[\begin{array}{c}
\nu_L^{}\\
e_L^{}
\end{array}\right]\,,
\end{array}
\begin{array}{l}q_{R}^{}(\textbf{3},\textbf{1},\textbf{2},~\,\frac{1}{3})=\left[\begin{array}{c}
u_R^{}\\
d_R^{}
\end{array}\right]\,,\\
[5mm]
l_{R}^{}(\textbf{1},\textbf{2},\textbf{1},-1)=\left[\begin{array}{c}
\nu_R^{}\\
e_R^{}
\end{array}\right]\,.\end{array}
\end{eqnarray}
As the left-right symmetry is the charge-conjugation, the scalars
and fermions will transform as
\begin{eqnarray}
\label{lrsymmetry}
\begin{array}{ccc}
\phi_{1,2}^{}\leftrightarrow\phi_{1,2}^{T}\,,&\quad\chi_{L}^{}\leftrightarrow\chi_{R}^{\ast}\,,&
\sigma\leftrightarrow \sigma\,,\\
[2mm] q_{L}^{}\leftrightarrow q_{R}^{c}\,,&\quad
l_{L}^{}\leftrightarrow l_{R}^{c}\,,&S_{R}^{}\leftrightarrow
S_{R}^{}\,.
\end{array}
\end{eqnarray}
We further impose a global symmetry, under which the fields carry
the quantum numbers as below
\begin{eqnarray}
\label{pqsymmetry}
\begin{array}{ccl}
1&\textrm{for}&q_L^{}\,,q_R^c\,,l_L^{}\,,l_R^c\,,S_R^{}\,;\\
[2mm]
2 &\textrm{for}&\phi_{1}^{}\,,\phi_{2}^{}\,,\sigma^\ast\,;\\
[2mm] 0 &\textrm{for}& \chi_{L}^{}\,,\chi_{R}^{}\,.
\end{array}
\end{eqnarray}
The full scalar potential then should be
\begin{eqnarray}
\label{potential}
V&=&\mu_\sigma^2|\sigma|^2_{}+\mu_\chi^2(|\chi_L^{}|^2_{}+|\chi_R^{}|^2_{})+\mu_{ij}^2\textrm{Tr}(\phi_i^\dagger\phi_j^{})\nonumber\\
&&
+\lambda_\sigma^{}|\sigma|^4_{}+\lambda_\chi^{}(|\chi_L^{}|^4_{}+|\chi_R^{}|^4_{})
+\lambda'^{}_\chi|\chi_L^{}|^2_{}|\chi_R^{}|^2_{}
\nonumber\\
&&
+\lambda_{ijkl}^{}\textrm{Tr}(\phi_i^\dagger\phi_j^{})\textrm{Tr}(\phi_k^\dagger\phi_l^{})
+\lambda'^{}_{ijkl}\textrm{Tr}(\phi_i^\dagger\widetilde{\phi}_j^{})\textrm{Tr}(\widetilde{\phi}_k^\dagger\phi_l^{})
\nonumber\\
&& +\kappa_{\sigma\chi}^{}|\sigma|^2_{}(|\chi_L^{}|^2_{}
+|\chi_R^{}|^2_{})+\rho_{ij}^{}|\sigma|^2_{}\textrm{Tr}(\phi_i^\dagger\phi_j^{})\nonumber\\
&&
+\xi^{}_{ij}(|\chi_L^{}|^2_{}+|\chi_R^{}|^2_{})\textrm{Tr}(\phi_i^\dagger\phi_j^{})\nonumber\\
&&+\xi'^{}_{ij}(\chi_L^\dagger\phi^{}_{i}\phi_j^{\dagger}\chi_L^{}+\chi_R^{T}\phi_i^T\phi_j^{\ast}\chi_R^\ast)
+[\alpha_{ij}^{}\sigma^2_{}\textrm{Tr}(\widetilde{\phi}_i^\dagger\phi_j^{})\nonumber\\
&&
+\beta_{i}^{}\sigma\chi_L^\dagger\phi_i^{}\chi_R^{}+\gamma_{i}^{}\sigma^\ast_{}\chi_L^\dagger\widetilde{\phi}_i^{}\chi_R^{}+\textrm{H.c.}]\,.
\end{eqnarray}
We also give the allowed Yukawa interactions,
\begin{eqnarray}
\label{yukawa} \mathcal{L}_Y^{}&=& -y_q^i
\bar{q}_L^{}\phi_i^{}q_R^{}-y_l^i
\bar{l}_L^{}\phi_i^{}l_R^{}-h(\bar{l}_L^{}\chi_L^{}S_{R}^{}+\bar{l}_R^{c}\chi_R^{\ast}S_{R}^{})\nonumber\\
&& -\frac{1}{2}g \sigma \overline{S}_R^{} S_R^{c} +\textrm{H.c.}\,.
\end{eqnarray}
Clearly, the lepton and quark doublets, the Higgs doublets, the
Higgs bi-doublets, the fermion singlets and the scalar singlet can,
respectively, belong to the $\textbf{16}_{F_i^{}}^{}$,
$\textbf{16}_H^{}$, $\textbf{10}_{H_{1,2}^{}}^{}$,
$\textbf{1}_{F_i^{}}^{}$ and $\textbf{1}_{H}^{}$ representation in
the $SO(10)$ GUTs.

The symmetry breaking pattern is expected to be
\begin{eqnarray}
\label{symmetrybreaking} &SU(3)_c^{}\times SU(2)_{L}^{}\times
SU(2)_{R}^{}\times U(1)_{B-L}^{}\times
U(1)_{\textrm{PQ}}^{}&\nonumber\\
&\downarrow\langle\sigma\rangle&\nonumber\\
&SU(3)_c^{}\times SU(2)_{L}^{}\times SU(2)_{R}^{}\times
U(1)_{B-L}^{}&\nonumber\\
&~\,\downarrow\langle\chi_R^{}\rangle&\nonumber\\
& SU(3)_c^{}\times SU(2)_{L}^{}\times
U(1)_{Y}^{}\nonumber\\
&~~\,\downarrow\langle\phi_{1,2}^{}\rangle&\nonumber\\
&SU(3)_c^{}\times U(1)_{em}^{}\,.&
\end{eqnarray}
We denote the VEVs by
\begin{eqnarray}
\begin{array}{c}
\langle\sigma\rangle=\frac{f}{\sqrt{2}}\,,~~\langle\chi_R^{}\rangle=\left[\begin{array}{c}\frac{v_R^{}}{\sqrt{2}}\\
[2mm] 0\end{array}\right]\,,~~\langle\chi_L^{}\rangle=\left[\begin{array}{c}\frac{v_L^{}}{\sqrt{2}}\\
[2mm] 0\end{array}\right]\,,\\
[5mm]
\langle\phi_{1}^{}\rangle=\left[\begin{array}{cc}\frac{v_{11}^{}}{\sqrt{2}}&0\\
[2mm]0&\frac{v_{12}^{}}{\sqrt{2}}\end{array}\right]\,,~~\langle\phi_{2}^{}\rangle=\left[\begin{array}{cc}\frac{v_{21}^{}}{\sqrt{2}}&0\\
[2mm]0&\frac{v_{22}^{}}{\sqrt{2}}\end{array}\right]\,,
\end{array}
\end{eqnarray}
which can be derived from the potential (\ref{potential}). Note that
$v_{11}^{}$, $v_{12}^{}$, $v_{21}^{}$, $v_{22}^{}$ and $v_{L}^{}$
should fulfill
\begin{eqnarray}
v=\sqrt{v_{11}^2+v_{12}^2+v_{21}^2+v_{22}^2+v_{L}^2}\simeq
246\,\textrm{GeV}\,.
\end{eqnarray}

\section{Peccei-Quinn symmetry}

After the complex scalar singlet $\sigma$ develops a VEV to
spontaneously break the global symmetry, it can be described by
\begin{eqnarray}
\sigma=\frac{1}{\sqrt{2}}(f+\rho)\exp\left(i\frac{a}{f}\right)\,.
\end{eqnarray}
Here $\rho$ is the physical boson while $a$ is the Nambu-Goldstone
boson (NGB). In the presence of the $\alpha,\beta,\gamma$-terms in
the potential (\ref{potential}), like the structure of the DFSZ
\cite{dfs1981} model, the NGB $a$ can couple to the quarks,
\begin{eqnarray}
\mathcal{L}\supset
-\frac{1}{2f}(\partial_\mu^{}a)\sum_{q}^{}\bar{q}\gamma^\mu_{}\gamma_5^{}q\,.
\end{eqnarray}
Therefore, through the color anomaly \cite{adler1969}, the NGB $a$
can pick up a tiny mass \cite{weinberg1978,bt1978},
\begin{eqnarray}
\label{amass1}
m_a^{2}=N^2_{}\frac{Z}{(1+Z)^2_{}}\frac{f_\pi^{2}}{f^2_{}}m_\pi^{2}\,,
\end{eqnarray}
where $N=3$ for three families of the SM quarks while $Z\simeq
m_u/m_d$. Clearly, the global symmetry is the PQ symmetry while the
NGB becomes a pseudo NGB (pNGB)---the axion. We can conveniently
express the axion mass as
\begin{eqnarray}
m_{a}^{}&=&\frac{\sqrt{Z}}{(1+Z)}\frac{f_\pi^{}}{f_{a}^{}}m_\pi^{}\simeq
6.2\,\mu\textrm{eV}\left(\frac{10^{12}_{}\,\textrm{GeV}}{f_a^{}}\right)\,,
\end{eqnarray}
where $f_a^{}$ is the axion decay constant. The PQ symmetry should
be broken at a high scale to fulfill the theoretical and
experimental constraints \cite{sikivie2008}. For example, the PQ
symmetry breaking may happen before the inflation \cite{guth1981} to
avoid the cosmological domain wall problem \cite{sikivie1982}. With
an appropriate $f_a^{}$ the axion can act as the dark matter
\cite{pww1983,hhrw2009}.

\section{Double and linear seesaw}

After the symmetry breaking (\ref{symmetrybreaking}), the fermions
will obtain their mass terms by their Yukawa couplings.
Specifically, the quarks and charged leptons can have the usual
$3\times 3$ Dirac mass matrices, i.e.
\begin{eqnarray}
\mathcal{L}\supset - \tilde{m}_u^{} \bar{u}_L^{}u_R^{} -
\tilde{m}_d^{} \bar{d}_L^{}d_R^{} - \tilde{m}_e^{}
\bar{e}_L^{}e_R^{}+\textrm{H.c.}\,,
\end{eqnarray}
with
\begin{eqnarray}
\begin{array}{c}
 \tilde{m}_u^{}=\frac{1}{\sqrt{2}}y^1_q v_{11}^{}+\frac{1}{\sqrt{2}}y^2_q v_{21}\,,\\
 [2mm]
 \tilde{m}_d^{}=\frac{1}{\sqrt{2}}y^1_q v_{12}^{}+\frac{1}{\sqrt{2}}y^2_q v_{22}\,,\\
 [2mm]
 \tilde{m}_e^{}=\frac{1}{\sqrt{2}}y^1_l v_{12}^{}+\frac{1}{\sqrt{2}}y^2_l v_{22}\,.
\end{array}
\end{eqnarray}
As for the neutral fermions including the left- and right-handed
neutrinos and the fermion singlets, they will form a symmetric
$9\times 9$ mass matrix as below,
\begin{eqnarray}
\label{lepton2} \mathcal{L}&\supset &
-\tilde{m}_\nu^{}\bar{\nu}_L^{}\nu_R^{}-\frac{1}{\sqrt{2}}hv_L^{}\bar{\nu}_L^{}S_R^{}
-\frac{1}{\sqrt{2}}hv_R^{}\bar{\nu}_R^c S_R^{}\nonumber\\
&&
-\frac{1}{2}M_S^{}\overline{S}_R^c S_R^{}+\textrm{H.c.}\,,\nonumber\\
&=&-\frac{1}{2}\left[\begin{array}{c}
\bar{\nu}_L^{} \\
[2.5mm]\bar{\nu}_R^c \\ [2.5mm] \overline{S}_R^c
\end{array}\right]^T_{}\left[\begin{array}{ccc}
0 ~~& ~~\tilde{m}_\nu^{} & \frac{hv_L^{}}{\sqrt{2}}\\
[2.5mm]
\tilde{m}_\nu^T & 0 & \frac{h v_R^{}}{\sqrt{2}}\\
[2.5mm] \frac{h^T_{} v_L^{}}{\sqrt{2}} &
\frac{h^T_{}v_R^{}}{\sqrt{2}}
 & M_S
\end{array}\right]\left[\begin{array}{c}
\nu_L^c \\
[2.5mm]\nu_R^{} \\ [2.5mm] S_R^{}
\end{array}\right]\nonumber\\
&&+\textrm{H.c.}\,,
\end{eqnarray}
with
\begin{eqnarray}
\tilde{m}_\nu^{}=\frac{1}{\sqrt{2}}y^1_l
v_{11}^{}+\frac{1}{\sqrt{2}}y^2_l v_{21}\,,~~
M_S^{}=\frac{1}{\sqrt{2}}gf\,.
\end{eqnarray}

\begin{figure*}
\vspace{3.8cm} \epsfig{file=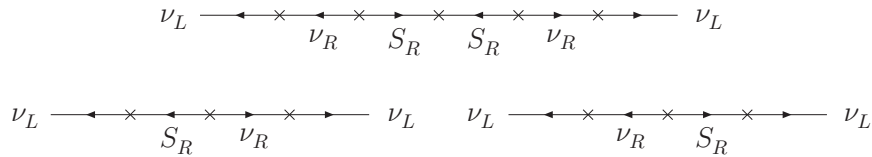, bbllx=5.3cm, bblly=6.0cm,
bburx=15.3cm, bbury=16cm, width=7.5cm, height=7.5cm, angle=0,
clip=0} \vspace{-8.5cm} \caption{\label{seesaw} Double seesaw (top)
and linear seesaw (bottom).}
\end{figure*}

For $\frac{1}{\sqrt{2}}hv_R^{}$ and/or $M_S^{}$ much bigger than
$\tilde{m}_\nu^{}$ and $\frac{1}{\sqrt{2}}hv_L^{}$, we can make use
of the seesaw formula \cite{minkowski1977} to derive the neutrino
masses,
\begin{eqnarray}
\mathcal{L}\supset
-\frac{1}{2}m_\nu^{}\bar{\nu}_L^{}\nu_L^c+\textrm{H.c.}\,,
\end{eqnarray}
where the mass matrix $m_\nu^{}$ contains two parts,
\begin{eqnarray}
\label{neutrino}
m_\nu^{}&=&\tilde{m}_\nu^{}\frac{1}{\frac{1}{\sqrt{2}}h^T_{}v_R^{}}M_S^{}\frac{1}{\frac{1}{\sqrt{2}}hv_R^{}}\tilde{m}_\nu^T
-(\tilde{m}_\nu^{}+\tilde{m}_\nu^T)\frac{v_L^{}}{v_R^{}}\,.\nonumber\\
&&
\end{eqnarray}
The first term is the double seesaw \cite{mohapatra1986,mv1986} for
$M_S^{}\gg \frac{1}{\sqrt{2}}hv_R^{}$ or the inverse seesaw
\cite{mv1986} for $M_S^{}\ll \frac{1}{\sqrt{2}}hv_R^{}$. Clearly,
the double seesaw should be our choice because of the large PQ
symmetry breaking scale. As for the second term, it is the linear
seesaw \cite{barr2003}. We can understand the double and linear
seesaw by Fig. \ref{seesaw}. In the double and linear seesaw
context, the leptogenesis \cite{fy1986} can be realized by the
decays of the right-handed neutrinos \cite{gs2010}.

\section{Summary}

In this paper we embedded the PQ symmetry in a left-right symmetric
theory and then in an $SO(10)$ GUT. In our model, the left- and
right-handed fermion doublets can naturally carry equal but opposite
PQ charges because the left-right symmetry is the
charge-conjugation. This implies we need two Higgs bi-doublets with
the PQ charge so that the fermions can obtain a desired mass
spectrum. The PQ symmetry breaking is driven by a complex scalar
singlet at a high scale. The invisible axion can pick up a tiny mass
through the color anomaly. In our model, the PQ symmetry breaking is
also responsible for generating the heavy Majorana masses of the
fermion singlets to realize the double and linear seesaw. Therefore,
the neutrino mass-generation is naturally related to the PQ symmetry
breaking \cite{shin1987}.

\textbf{Acknowledgement}: I thank Manfred Lindner for helpful
discussions. This work is supported by the Alexander von Humboldt
Foundation.

\end{document}